\documentclass[12pt,preprint]{aastex}

\shorttitle{NGC~346 PDMF}
\shortauthors{Sabbi et al.}

\begin{document}

\title{THE STELLAR MASS DISTRIBUTION IN THE GIANT STAR FORMING REGION NGC~346\footnote{Based on observations with the NASA/ESA Hubble Space
Telescope, obtained at the Space Telescope Science Institute, which is
operated by AURA Inc., under NASA contract NAS 5-26555}}

\author{E. Sabbi\altaffilmark{2}, M. Sirianni\altaffilmark{2,3}, A.
Nota\altaffilmark{2,3}, M. Tosi\altaffilmark{4},  J.
Gallagher\altaffilmark{5}, L.J. Smith\altaffilmark{2,3,6}, L.
Angeretti\altaffilmark{4}, M. Meixner\altaffilmark{2}, M.S.
Oey\altaffilmark{7}, R. Walterbos\altaffilmark{8}, \& A.
Pasquali\altaffilmark{9}}
\email{sabbi@stsci.edu}

\altaffiltext{2}{Space Telescope Science Institute, 3700 San Martin Drive,
Baltimore, USA}
\altaffiltext{3}{European Space Agency, Research and Scientific Support Department, Baltimore, USA}
\altaffiltext{4}{INAF--Osservatorio di Bologna, Italy}
\altaffiltext{5}{University of Wisconsin, Madison, WI, USA}
\altaffiltext{6}{University College London, London, UK}
\altaffiltext{7}{University of Michigan, Ann Arbor, MI, USA}
\altaffiltext{8}{New Mexico State University, Las Cruces, NM, USA}
\altaffiltext{9}{Max-Plank-Institut f\"{u}r Astronomie, Heidelberg, Germany}

\begin{abstract}

Deep F555W and F814W {\it Hubble Space Telescope\/} ACS images are the basis for a study of the present day mass function (PDMF) of NGC~346, the largest active star forming region in the Small Magellanic Cloud (SMC). We find a PDMF slope of $\Gamma=-1.43\pm 0.18$ in the mass range $0.8-60\, {\rm M}_\odot$, in excellent agreement with the Salpeter Initial Mass Function (IMF) in the solar neighborhood. Caveats on the conversion of the PDMF to the IMF are discussed.  The PDMF slope changes, as a function of the radial distance from the center of the NGC~346 star cluster, indicating a segregation of the most massive stars. This segregation is likely primordial considering the young age ($\sim 3\, {\rm Myr}$) of NGC~346, and its clumpy structure which suggests that the cluster has likely not had sufficient time to relax.
Comparing our results for NGC~346 with those derived for other star clusters in the SMC and the Milky Way (MW), we conclude that, while the star formation process might depend on the local cloud conditions, the IMF does not seem to be affected by general environmental effects such as galaxy type, metallicity, and dust content.
\end{abstract}


\section{INTRODUCTION}
\label{intro}

Since the pioneering study by \citet{salpeter55}, one of the major issues in astrophysics is whether the initial mass function (IMF), the relationship that specifies the mass distribution of a newly formed stellar population, is  universal or, alternatively, determined by environmental effects. In particular, establishing whether the IMF has been constant over the evolution of the universe, or if it varies with time (redshift) and/or metallicity has crucial consequences on the evolution, surface brightness, chemical enrichment and baryonic content of galaxies, and on the evolution of light and matter in the universe. 

Due to a number of effects and observational uncertainties such as mass segregation, field star contamination, and variable extinction \citep{elmegreen06}, the ``true" IMF is very difficult to measure. Despite these limitations, star clusters still provide the best determination of the IMF, because they are essentially single--age, single--metallicity stellar systems. In general, all star clusters lose more than 80\% of their stellar content in their first $\sim 10$ Myr \citep{kroupa02,lada03}. Therefore, in order to perform an accurate census of the initial stellar content it is necessary to focus on the youngest stellar systems. In addition, young (age $<3-5\, {\rm Myr}$) star clusters in nearby galaxies (e.g. the Large --LMC-- and Small Magellanic Cloud --SMC) offer to us the unique advantage to individually resolve --- because of their proximity --- statistically significant samples of stars down to the sub--solar mass regime. 

In this paper, we focus our attention on the mass function (MF) of NGC 346, an extremely young \citep[$\sim 3\, {\rm Myr}$,][hereafter S07]{bouret03,sabbi07}, and active star cluster that excites the largest and brightest H{\sc ii} region (N66) in the SMC.   NGC~346 ($\alpha_{j2000}=00^h 59^m 05.2^s, \delta_{j2000}=-72^{\deg} 10'28''$) contains a major fraction of the O stars known in the entire SMC \citep{walborn78,walborn86,niemela86,massey89}. The bright end of its stellar population ($V\le 19.5\, {\rm mag}$) has been well investigated in the past 20 years. Spectral investigations of the brightest members identified several stars of spectral type O6.5 or earlier \citep{massey89,heap06,mokiem06,evans06}. \citet{massey89} found the  MF for  the brightest stars in NGC 346 --- down to 5 M$_{\odot}$ --- to have a slope that is consistent with what has been found for massive stars near the Sun and in the LMC. 

High resolution observations obtained with the Advanced Camera for Survey (ACS) on board of the \emph{Hubble Space Telescope} (\emph{HST}) recently revealed that the star formation history of the region where NGC~346 is located is quite complex. There is evidence for old episodes of star formation (SF), in the field of the SMC approximately between 3 and 5 Gyr ago. After that, the SF activity in this region appears to have been significantly diminished, with a possible moderate enhancement $\sim 150\, {\rm Myr}$ ago (S07). An intermediate--age ($\sim 4.5\, {\rm Gyr}$) stellar cluster (BS90), characterized by a core radius $r_c\simeq 25\, {\rm arcsec}$ and a tidal radius $r_t \simeq 130\, {\rm arcsec}$, is located at a projected distance of $\sim 23\, {\rm pc}$ northeast from the center of NGC~346 (S07). Spectral analysis of NGC~346 members revealed the presence of stars as young as $\sim 1\, {\rm Myr}$, \citep{massey05}, but stars $\sim 5\, {\rm Myr}$ old are also present \citep{heap06,mokiem06}.

Star formation is likely still active in the region: hundreds of pre--main sequence (pre--MS) stars in the mass range $0.6 - 3\, {\rm M}_{\odot}$ were discovered by \citet{nota06} from \emph{HST}/ACS images. \emph{Spitzer Space Telescope} (\emph{SST}) observations also detected a myriad of embedded young stellar objects (YSOs) scattered across the entire region \citep{bolatto07}. By assuming a Salpeter IMF, \citet{simon07} calculated that more than 3000 M$_\odot$ have been formed in the last $\sim 10^6\, {\rm yr}$, concluding that $\ge 6$\% of the current SF in the SMC is taking place in NGC~346. 

This recent stellar population does not appear to be uniformly distributed within the ionized nebula, but is rather organized in many, likely coeval, subclusters (S07) that are coincident with clumps of molecular gas previously detected by \citet{rubio00}. \citet{simon07} noted that almost all the subclusters host one or more YSOs.

The high sensitivity of the \emph{HST}/ACS allowed us to investigate the stellar content of NGC~346 from $\sim 60 M_\odot$ down to $\sim 0.6 M_\odot$, making NGC~346 one of the few known regions where a MF can be determined over two orders of magnitude \citep[the other classical case being the Orion trapezium cluster in our own MW;][]{elmegreen06}. This paper gives a new derivation of the NGC 346 MF, reviews the strengths and limitations of the methods, and presents the conclusions that we can conservatively derive. The paper is organized as follows: a short description of the data reduction procedure is presented in \S\ref{obs}; in \S\ref{CMD} we present
the color--magnitude diagram (CMD), and we describe the stellar populations identified. In \S\ref{MF} we present the MF we obtained for the NGC~346 cluster, we discuss its spatial variations, and we analyze the impact of the environment on the MF. The results of this paper are discussed in \S\ref{conclusions}.

\section{OBSERVATIONS AND DATA REDUCTION}
\label{obs}

Multiple, deep, images of NGC~346 were obtained during Cycle 14 using the Wide Field Channel (WFC) of the \emph{HST}/ACS (GO-10248; PI A. Nota). A detailed description of the observations and of the stellar photometry has been presented in S07. Several exposures were taken through the filters F555W ($\sim$V) and F814W ($\sim$I). A dither pattern was especially designed to allow for hot pixel removal, improve the sampling of the point spread function and the accuracy of the photometry by averaging the flat-field errors and by smoothing over the spatial variations in the detector response.
Short and long exposures were taken to maximize the photometric range.

The entire dataset was processed adopting the standard STScI ACS calibration pipeline (CALACS), and exposures in each band were co--added using the MULTIDRIZZLE package \citep{koekemoer02} to correct for the geometrical distortions, and remove cosmic rays and hot pixels. The final exposure time of the deep combined F555W and F814W images is $\simeq 4100\, {\rm s}$, and the total area imaged is $\simeq 5' \times 5'$, corresponding to $\simeq 88\times 88\, {\rm pc}$, assuming a distance to the SMC of 60.6 Kpc \citep{hilditch05}. The photometric reduction of the images was performed with the DAOPHOT package within the IRAF\footnote{IRAF is distributed by the National Optical Astronomy Observatory, which is operated by AURA Inc., under cooperative agreement with the National Science Foundation.} environment. We followed the same methodology applied to R136 \citep{sirianni00} and NGC~330 \citep{sirianni02} and used the PSF fitting and aperture photometry routines provided within DAOPHOT to derive accurate photometry of all stars in the field (see S07 for more details).

The instrumental magnitudes were corrected for the charge transfer efficiency (CTE) effects that affect ACS data, using the parametric correction from \citet{reiss04}. The photometric calibration to the Vegamag system was performed by converting the magnitudes of the individual stars to an aperture radius of 0.5$''$ and applying the zero points listed in \citet{sirianni05}.

\begin{figure}
\epsscale{1.}
\plotone{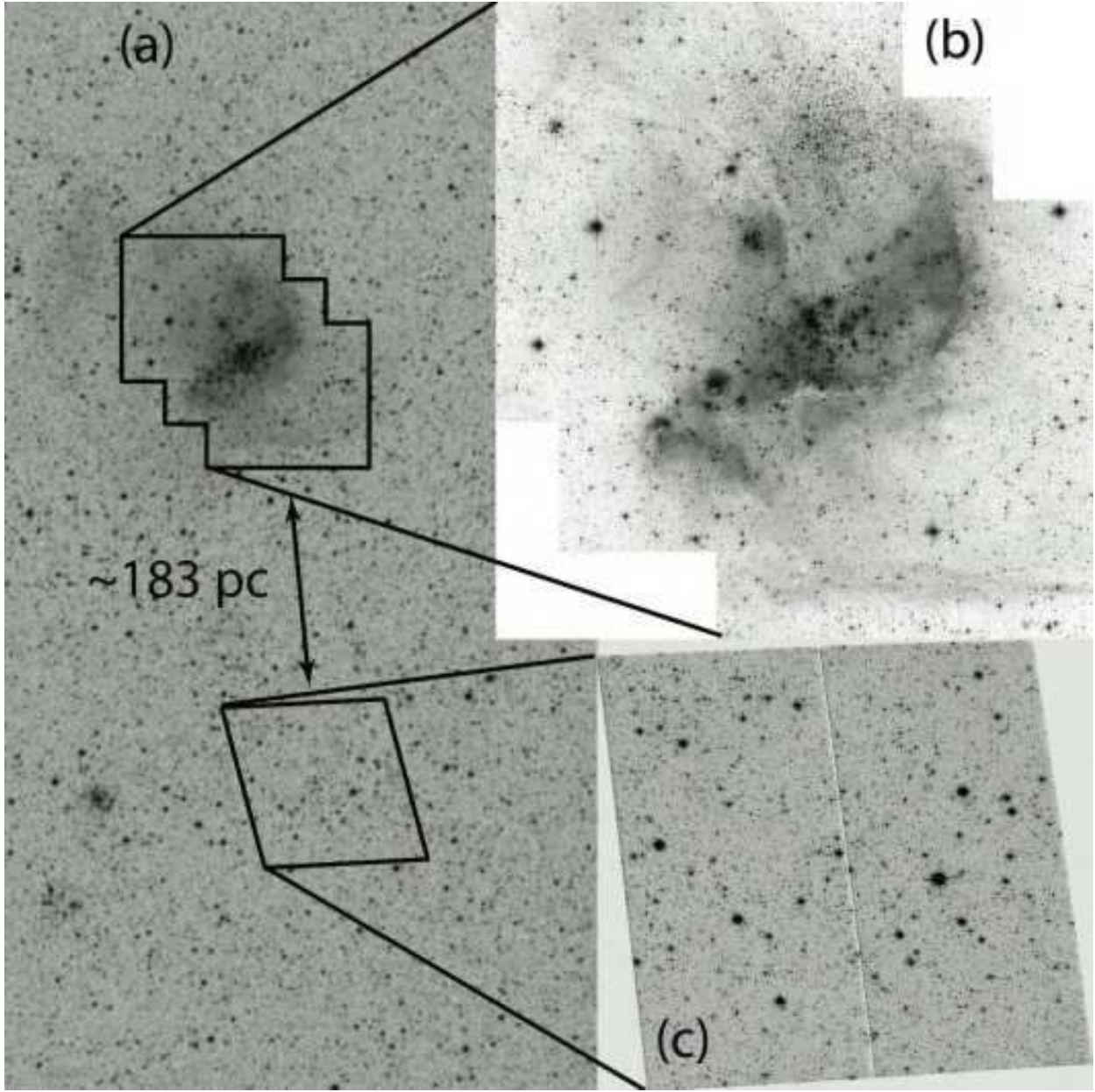}
\caption{\label{f:mappa_field} Relative position of NGC~346 and of the field observed as reference (a). An enlarged image of NGC~346 is shown in panel (b), together with an enlarged view of the field~(c).}
\end{figure}

A SMC field ($\alpha=$ 00:58:42.5; $\delta = -$72:19:46) was observed with WFC for comparison. The observed field is located $\sim9' 16''$ south ($\sim 163\, {\rm pc}$ in projection) from NGC~346 (Fig.~\ref{f:mappa_field}). The data were analyzed following the same procedure described above.

Extensive artificial star tests were designed to accurately assess the level of completeness of the photometric data. More than 3,000,000 artificial stars --- obtained from the scaled PSF used for the photometric analysis of the frames --- were added to both the F555W and the F814W images using the subroutine {\it addstar\/} in DAOPHOT, and subsequently retrieved. 
The artificial stars were distributed onto the F555W image randomly in magnitude using a step function. This function extends two magnitudes below the detection limit of our observations in order to probe with sufficient statistics the range of magnitude in which incompleteness is expected to be most severe. This is a crucial step to assess the robustness of the MF.

In each run $\sim 7000$ artificial stars were distributed on a regular grid to avoid an unrealistic increase of the crowding conditions in the analyzed frame. Then, the frame to which the artificial stars were added was processed exactly as the original frame. We determined the level of completeness of the photometry by comparing the list of artificial stars added with the list of the stars recovered.
In our experiments an artificial star is considered recovered if it is found in both F555W and F814W frames with an input-output difference in magnitude $\Delta m \le 0.75$, and at the same time satisfying all the photometric selection criteria. 

The artificial star experiments show that the completeness is a function of both magnitude and color: our photometry is complete at a 50\% level down to magnitude $m_{\rm F814W} \simeq 25$ at color $m_{\rm F555W}-m_{\rm F814W} \simeq -0.2$. At redder colors (e.g. $m_{\rm F555W}-m_{\rm F814W} \simeq 2.0$) the photometry is complete at the 50\% level only at magnitude $m_{\rm F814W} \simeq 23.2$ (Fig.~\ref{f:cmd}). The 50\% level of completeness defines the magnitude limit we have conservatively imposed for the MF determination. 

\section{THE STELLAR POPULATIONS IN NGC~346}
\label{CMD}

Figure~\ref{f:cmd} shows the $m_{\rm F814W}$ vs. $m_{\rm F555W}-m_{\rm F814W}$ CMD of all the stars identified in the WFC NGC~346 data with $\sigma_{DAO}<0.1$ mag. We can distinguish in this CMD the different stellar populations that coexist in the region.

$\bullet$ A rich old stellar population, with a  main sequence (MS) extending from $m _{\rm F814W}\simeq 22.0$ down to $m _{\rm F814W}\simeq 25.5$. The evolved stars of this population clearly define the red giant branch (RGB), with the brightest stars at $m _{\rm F814W}\simeq 15.0$, and a tight red clump (RC) at $m _{\rm F814W}\simeq 18.5$. The narrow subgiant branch (SGB), visible at magnitude $m _{\rm F814W}\simeq 21.0$, in the color range $0.45\le m _{\rm F555W}-m _{\rm F814W}\le 0.95$, indicates that, in the NGC~346 region, the old stellar population is dominated by the presence of the BS90 star cluster. The presence of SGB stars both fainter and brighter than the BS90 SGB indicates that, in the field, SF occurred before and after the formation of this star cluster. 

$\bullet$ A young population. The bright ($12.5\le m _{\rm F814W}\le 22.0$) and blue ($-0.3\le m _{\rm F555W}-m _{\rm F814W}\le 0.4$) MS, well visible to the upper left of the CMD, is the most remarkable feature of the youngest stellar population. The majority of these stars belong to the cluster NGC~346, and have an age of $\sim 3\, {\rm Myr}$. 

$\bullet$ Hundreds of pre-MS stars are visible between $0.6\le m _{\rm F555W}-m _{\rm F814W}\le 2.2$, below magnitude $m _{\rm F814W}\simeq 22.0$. 

\section{THE MASS FUNCTION}
\label{MF}

The MF, commonly indicated by $\xi$, is defined as the fractional number of stars per mass interval per unit area. Among the various parameterizations of the MF \citep[see][]{kroupa02}, one of the most used is the one proposed by \citet{scalo86}, where the MF is characterized by the logarithmic derivative $\Gamma = d(\log\xi(\log(m))/d\log m$, where $\xi(\log(m)$ is the MF, and $\Gamma$ is its slope. The reference IMF, as derived by \citet{salpeter55} for the solar neighborhood, has a slope $\Gamma = -1.35$. 

The MF of a system is obtained by counting the number of stars in mass intervals. Under the assumption that all the stars of a system formed at the same time, the MF can be derived by counting the number of stars as a function of magnitude (luminosity function), which translates into the number of stars per unit stellar mass by using an appropriate mass-luminosity relation \citep[see e.g.][]{degrijs02}. However, this approach can be applied only to those stars currently on the MS, where the relationship between the luminosity of a star and its mass is reasonably well known, and assumes that all the stars have the same age, and the same metallicity, or, in other words, that we are studying a simple stellar population (SSP).

As discussed earlier, there is evidence that NGC~346 formed stars over the last 5 Myr. Furthermore hundreds of stars in NGC~346 are still in the pre--MS phase (where their luminosity strongly depends on age), and star formation is still ongoing \citep{bolatto07,simon07}. For all these reasons we cannot model NGC~346 as a perfect SSP. To overcome the age--luminosity degeneracy which comes from the fact that NGC~346 is not a SSP, we applied the technique presented by \citet{tarrab82} and \citet{massey95}, which consists of directly counting the stars between two theoretical evolutionary tracks, according to their position in the Herztprung--Russell Diagram (HRD). This technique has the advantage of avoiding any assumptions on the age of the stellar population. 

Because in NGC~346 SF occurred over an interval of time comparable with its age and is still ongoing \citep{simon07}, it is almost impossible to derive its IMF without making assumptions on how the star formation rate varies in time, and on the mass distribution of the stars which are currently forming. For this reason we will focus our attention on the present--day mass function (PDMF) of those stars that are already visible in the optical bands, but have not yet evolved off the MS.

\begin{figure}
\epsscale{1.}
\plotone{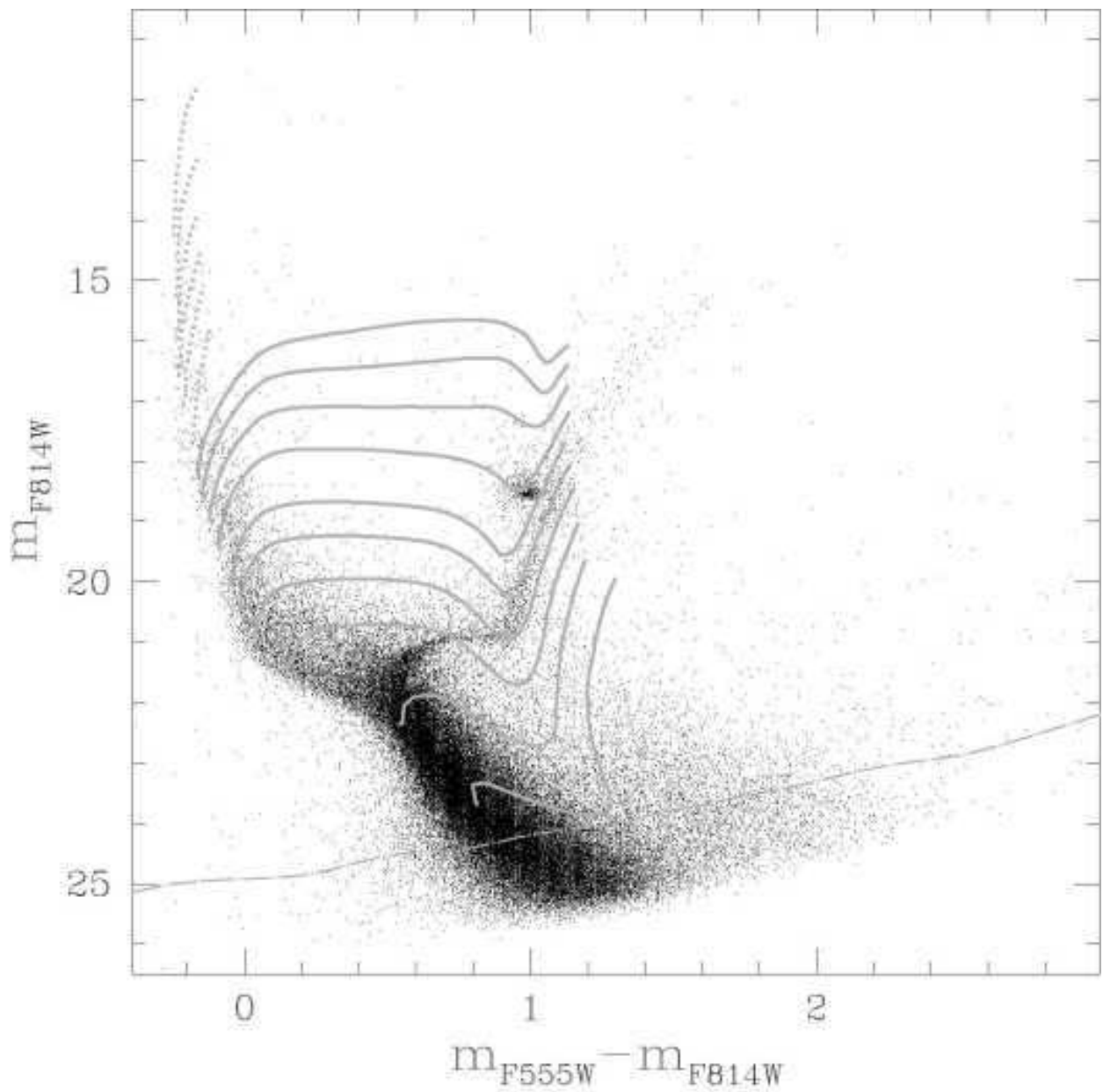}
\caption{\label{f:cmd} NGC~346 $m_{F814W}$ vs. $m_{F555W}-m_{F814W}$ CMD. We show Padua MS evolutionary tracks for stars more massive than 9 M$_\odot$ (grey dotted lines) and Siess Pre-MS evolutionary track for stars in the mass range between 0.8 and 7 M$_\odot$ (grey continuous lines). The grey dashed line indicates the 50\% level of completeness of our photometry, as derived from the artificial star experiments.}
\end{figure}

To infer the PDMF of NGC~346, we used two different sets of evolutionary tracks: for stars below 7 M$_\odot$, we adopted pre--MS stars evolutionary tracks by \citet{siess00}, while we used Padua evolutionary tracks \citep{fagotto94} for stars above 9 M$_\odot$. In order to minimize the uncertainties due to the photometric conversions, we transform the evolutionary tracks into the Vegamag system by applying the convertions calculated by \citet{origlia00}. The adopted evolutionary tracks, superimposed on the $m_{F555W}$ vs. $m_{F555W}-m_{F814W}$ CMD, are shown in Figure~\ref{f:cmd}.

We computed the PDMF between 0.8 and 60 M$_\odot$ using the following binning: 0.8, 1.2, 1.6, 2.0, 2.5, 3.0, 4.0, 5.0, 6.0, 7.0, 9.0, 12.0, 15.0, 20.0, 30.0, and 60.0 M$_\odot$.

\begin{figure}
\epsscale{1.}
\plotone{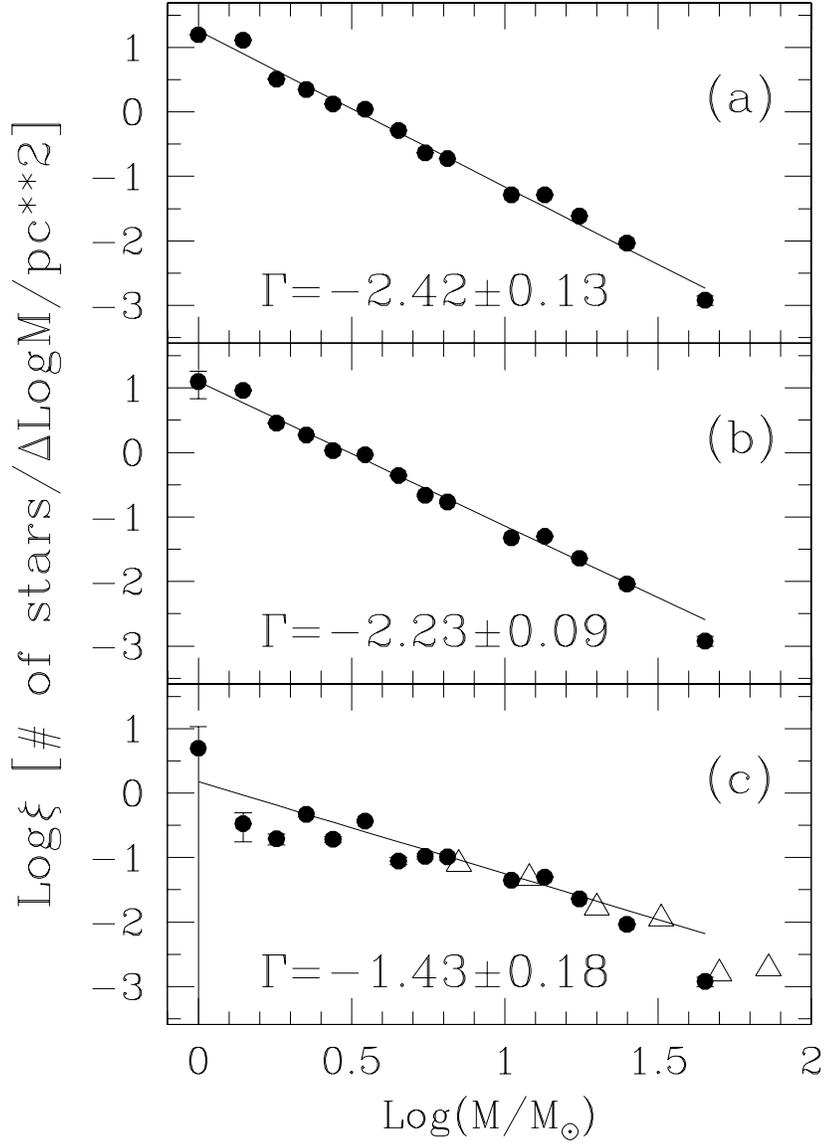}
\caption{\label{f:MFtot} PDMF of NGC~346 in the mass range between 0.8 and 60 M$_\odot$. Panel (a) shows the derived PDMF before removing the contributions from the old star cluster BS90 and the field of the SMC; panel (b) shows the PDMF after the correction for the cluster BS90, and panel (c) the ``true'' PDMF of NGC~346, after the subtraction of BS90 and the field of the SMC. Open triangles represent the IMF as derived by \citet{massey89}.}
\end{figure}

To properly account for the effects of crowding and incompleteness, which varies with both magnitude and color (e.g. the dashed line in Fig.~\ref{f:cmd} shows the 50\% level of completeness of our photometry), stars identified between two evolutionary tracks were divided into bins of 0.5 in magnitude and 0.25 in color. The number of stars identified in each bin was then normalized to the logarithmic width of the mass range spanned by the tracks and to the area of the observed region. We then applied the appropriate completeness factor. Figure~\ref{f:MFtot}--panel (a) shows the PDMF obtained with this approach.

The resulting PDMF reflects the composite stellar population in this region. In order to derive the "true" PDMF for the NGC 346 cluster, we need to account for the contamination of the old cluster BS90 and of the SMC field.

$\bullet$ {\it The BS90 old cluster}. We used the King model that best fits the stellar density profile of BS90 (S07) to determine, at each spatial position, the expected stellar contribution from BS90. We selected the stars that lie within 20 arcsec from the projected center of gravity of BS90 as representative of the old cluster stellar population. We then scaled the number of stars to the estimated stellar density, and removed them from our PDMF. We found that $\approx$11\% of the stars are likely to belong to BS90. Figure~\ref{f:MFtot}--panel (b) shows  the PDMF after the subtraction of the contribution from BS90.

$\bullet$ {\it The SMC field.} We used a SMC field ($\alpha_{J2000}=00^h58^m42.5^s, \delta_{J2000}=-72\degr 19'46''$), located at a projected distance of $\sim 163\, {\rm pc}$ from NGC~346, observed with WFC in the same filters, to remove the contribution of the SMC stellar population from our data. This contribution is important, given that the surface stellar density in this region is $\sim 9.8$ stars pc$^{-2}$. As we did for NGC~346, we counted the number of stars lying within two evolutionary tracks, normalized to the logarithmic width of the mass range spanned by the tracks and to the area of the observed region, and then we applied the appropriate completeness factor. Then for each mass bin, we subtracted the estimated number of stars belonging to the field from the PDMF of NGC~346. 
From the comparison between the reference field and NGC~346 region, we can conclude that in NGC~346 region about 54\% of the stars below 12 M$_\odot$ belong to the SMC field. 

The final PDMF of NGC~346, after the subtraction of the SMC field and BS90, is shown in Figure~\ref{f:MFtot}--panel (c). A weighted least mean square fits of the data indicates that, between 0.8 and 60 M$_\odot$, the slope of NGC~346 PDMF is $\Gamma=-1.43\pm 0.18$, in good agreement with the value derived by Salpeter for the IMF of the solar neighborhood. Furthermore we note that, as already derived by \citet{massey89}, the PDMF is quite steep above 5 M$_\odot$ (we derive a slope $\Gamma=1.87\pm 0.41$), and becomes flatter below this value of mass.

Some uncertainties still affect the derived slope of the PDMF:
\begin{itemize}
\item As \citet{massey95} have shown, optical and near--UV colors only sample the low--frequency tail of the spectral energy distribution of the most massive and hot stars; as a consequence, optical photometry is not adequate to estimate the PDMF of massive stars, and spectroscopy is necessary. In Figure \ref{f:MFtot}--(c) we have compared our derived PDMF with that published by \citet{massey89} (open triangles): we found a very good agreement in the overlapping range of masses below 20 M$_\odot$. As expected, on the basis of photometry alone, we underestimate the masses of the most massive stars.   
\item The presented PDMF does not take into account uncertainties due to unresolved binary systems, which may be important. In the Orion Nebula Cluster, for example, each massive star has, on average, 1.5 companions \citep{priebisch99}. \citet{sagar91} estimated that, if each star in the mass range 2 to 14 M$_\odot$ has one companion, the average IMF slope they derived for five young clusters in the LMC would change from $\Gamma=-1.3$ to $-1.7$. This correction, however, depends on the distribution of the companions, which is not yet well-known \citep{priebisch99,duchene01}. A spectroscopic survey of NGC~346, carried on with the Fiber Large Array Multi--Element Spectrograph (FLAMES) instrument at the Very Large Telescope (VLT), indicates a binary fraction of $\approx 26$\% for the most massive stars \citep[$5\le M \le 60\, {\rm M}_\odot$,][]{evans06}.
\item The derivation of the PDMF requires an accurate knowledge of the amount of extinction. The foreground extinction to the SMC is low, and the only appreciable presence of gas and dust  along the line of sight of NGC~346 is likely associated with the star--forming region itself. A visual inspection of the region shows that dust and gas are not uniformly distributed, but detailed reddening maps for the region are not yet available. 
\item The real slope of the MF can be altered by an inaccurate subtraction of contaminating field stars: dust and gas for example can obscure field stars, causing an over subtraction of the low--mass stars, and a flatter MF. Local differences in the stellar density or in the distance dispersion between the analyzed region and the reference field may also artificially alter the shape of the MF (see \S\ref{radialMF}).
\item A major source of uncertainty comes from the pre--MS evolutionary tracks, which do not yet well represent observations \citep[see for example the discussions in][]{chabrier03,hillenbrand07}.
\end{itemize}

\subsection{The spatial variations of the PDMF}
\label{radialMF}

The high spatial resolution and the depth of our photometric data have allowed us to study in detail the spatial variations of the PDMF within NGC~346. Already a cursory inspection of the image shows that the brightest stars are found in the central region. Does this imply that the most massive stars are concentrated in the cluster center?

To answer this question, we have calculated the variation of the mass function with projected distance from the NGC~346 center. In order to determine the position of the projected center of gravity of NGC~346 we first selected the stars belonging to the young MS (defined by magnitude $m_{\rm F814W}$ brighter than 21.0 and colors $m_{\rm F555W}- m_{\rm F814W}$ bluer than 0.4). 
Then we drew a circle with a radius of 50 pixel around each star and counted the number of stars falling within each circle. We computed the weighted average of the MS stars $\alpha$ and $\delta$ coordinates, using the sum of the stars within each circle as weight. The position we obtained for the stellar association is $\alpha = 00^h59^m05\farcs 8\, \delta = -72\degr 10^m 35\arcsec, [J2000]$, with a 1$\sigma$ uncertainty in both $\alpha$ and in $\delta$ of $\sim 1$, that corresponds to about 25 pixels in the HST/WFC images.

\begin{figure}
\epsscale{1.}
\plotone{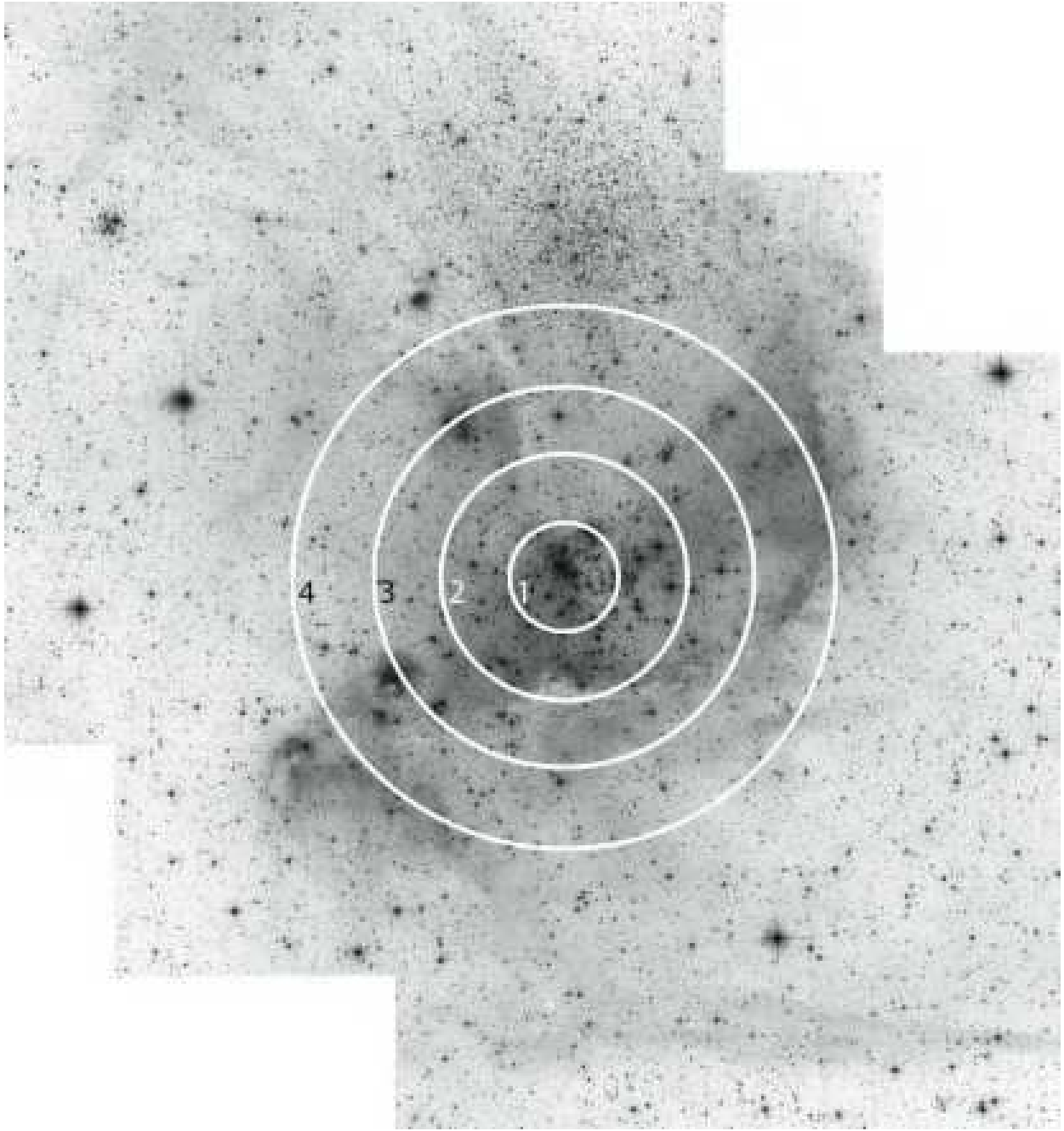}
\caption{\label{f:mappa_N346} NGC~346 WFC/ACS mosaiked data. We show the four annuli used to study the spatial variations of the PDMF.}
\end{figure}

\begin{figure}
\epsscale{1.}
\plotone{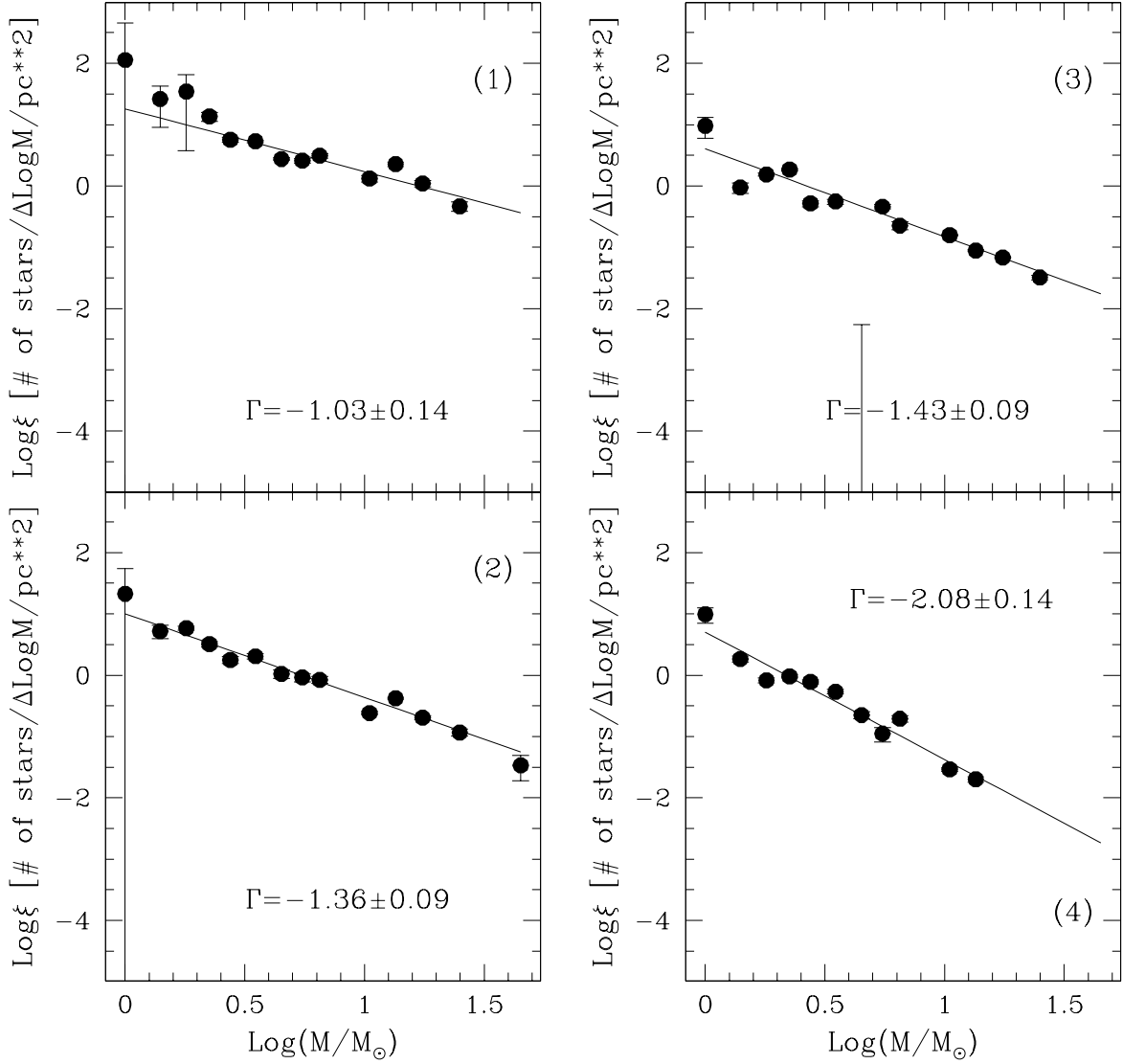}
\caption{\label{f:MFrad} PDMFs of the four regions of NGC~346 analyzed: in panel (1) we present the PDMF derived within 4 pc from the center of the cluster, panel (2) shows the PDMF obtained between 4 and 9 pc, the PDMF obtained between 9 and 14 pc is shown in panel (3), and in panel (4) we show the PDMF corresponding to 14  and 20 pc.}
\end{figure}

\begin{figure}
\epsscale{1.}
\plotone{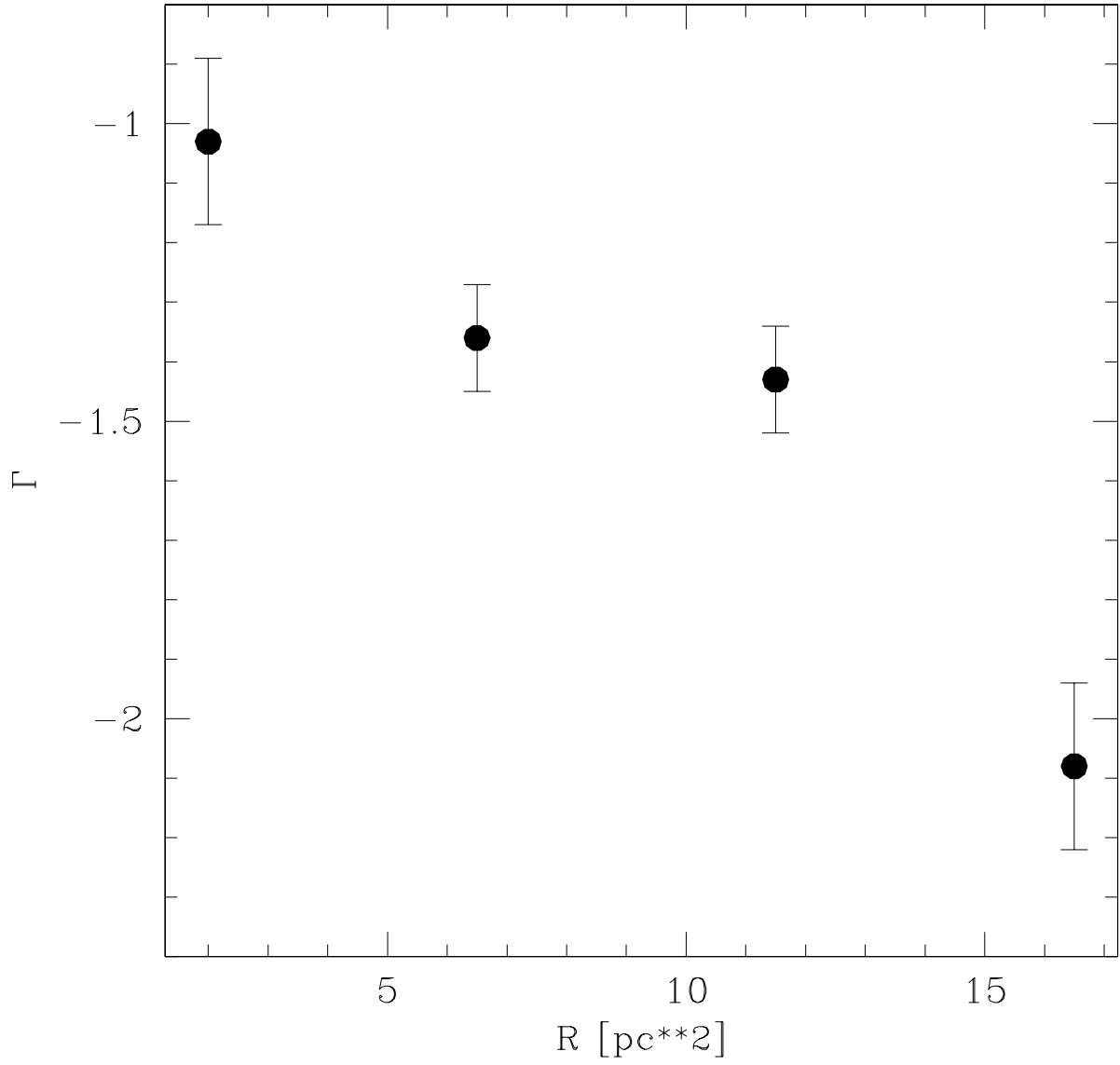}
\caption{\label{f:esp} Variation of the PDMF slope $\Gamma$ as a function of the radial distance from the center of NGC~346.}
\end{figure}

To study the spatial variations of the PDMF, we divided the region in four annuli, at radial distances from the center R$\le$ 4.00, 9.00 (which corresponds to the half mass radius), 14.00, and 19.80 pc (Fig.~\ref{f:mappa_N346}). Less than 2.6 stars/pc$^2$ belong to NGC~346 outside $\sim 20\, {\rm pc}$ from the center.  PDMFs derived for annuli 1, 2, 3, and 4 are shown in Figure~\ref{f:MFrad}, and the corresponding slopes, as a function of the distance from the center, are shown in Fig~\ref{f:esp}.  In the innermost region the PDMF appears quite flat, with $\Gamma=-1.03\pm 0.14$ (Fig~\ref{f:MFrad}). It becomes steeper moving from the center to the periphery, where its exponent becomes $\Gamma=-2.08\pm 0.14$. 

We applied a Kolmogorov--Smirnov (K--S) test to verify the significance of the differences found between the PDMF obtained in the innermost annulus (annulus \# 1) and those derived for the other three annuli. 
According to the K--S test, the probability that the PDMFs derived in annuli 2, 3, and 4 are drawn from the parent distribution of the PDMF in the innermost annulus is 3.9\%, 0.3\%, and 0.06\% respectively. Therefore, we can safely exclude that regions 3 and 4 are drawn from the same parent distribution of region 1.

The projected density of massive stars from the innermost annulus to the periphery decreases by a factor of 60, whereas low-mass stars are depleted only by a factor of 6, indicating that the change in the PDMF slope is due to a lack of massive stars in the periphery, rather than an excess of low mass stars there. A similar trend was observed in Orion Nebula Cluster (ONC).  
Discussing the excess of massive stars in the center of the ONC \citet{hillenbrand97} pointed out that mass distribution could be biased by an age effect, with the most massive stars also being the youngest. This does not seem to be the case for NGC~346: spectroscopic analysis by \citet{mokiem06} have demonstrated that, even if a spread of about 5 Myr in the ages of NGC~346 stars is present, stars of different ages appear uniformly distributed over the whole cluster, and SST infrared observations indicate that there is ongoing SF throughout the complex with a lower limit on the star formation rate of $3.2 \times 10^{-3}\, {\rm M}_\odot\, {\rm yr}^{-1}$ \citep{simon07}.

As already mentioned in \S\ref{MF}, dust and gas can obscure field stars, and this may cause an over subtraction of the low--mass stars, resulting in a flatter MF. To exclude the possibility that the spatial variation of the MF can be totally ascribed to variations in the gas and dust densities, we derived the MF of the innermost region also under the assumption that the contamination from the SMC field is negligible, and in this case we obtained a value for $\Gamma=-1.11\pm 0.13$. We also estimated the surface stellar density necessary to obtain the same MF slope in the innermost region that we find for the outermost one. In order to obtain a slope of $\Gamma=-2.08$ we should observe a stellar density of $\sim 1000$ star pc$^{-2}$ between $0.8 - 1.2\, {\rm M}_\odot$, which is almost a factor of 2 higher than the observed value; such a high stellar density makes it unlikely that the derived variation can be ascribed only to a poor subtraction of the SMC field.

\subsection{Is the mass segregation primordial?}
\label{prim-segr}

As for many other young star clusters, NGC~346 shows a shallower PDMF in its central region. It is still a matter of debate whether the segregation of the most massive stars in young clusters is due to evolutionary effects, or instead is due to the initial conditions of the cluster formation. In the first case the most massive stars form elsewhere in the cluster and then sink to the center through dynamical interactions with the numerous low-mass stars. On the contrary, if the mass segregation is due to initial conditions, then it reflects how the cluster formed.

As discussed by S07, NGC~346 shows a very complex morphology, with the stars assembled in many coeval sub--clusters. Due to the lack of information on the dynamics of the sub--clusters, at zero order we can consider all the stars in NGC~346 as members of an individual star cluster. Thus, to understand whether the mass segregation found in NGC~346 is primordial or not, we can compare its age with its mass segregation timescale ($T_\mathrm{msg}$), defined as $T_\mathrm{msg}=2 T_\mathrm{relax} m_\mathrm{av}/m_\mathrm{max}$, $T_\mathrm{relax}$ is the relaxation time of the cluster, $m_\mathrm{av}$ is the average stellar mass and $m_\mathrm{max}$ is the mass of the most massive star \citep{kroupa04}. For NGC~346 we found $T_{msg}\simeq 2\times 10^7$ yr (values used to derive the $T_\mathrm{msg}$ are listed in Table~\ref{table}), one order of magnitude larger than the age of NGC~346, supporting the idea that the observed segregation of the massive stars is likely due to initial conditions, rather than dynamical evolution.

It has been suggested that NGC~346 is probably the result of the collapse and subsequent fragmentation of the initial giant molecular cloud into multiple "seeds" of SF. The fact that the observed subclusters appear almost coeval, and that many of them appear to be connected by arcs of dust and gas, seems to be a good observational match to the conditions predicted by the hierarchical fragmentation of a turbulent molecular cloud model \citep{klessen00,bonnell02,bonnell03}.
If this is the case, the fact that the various ``seeds'' of star formation are still distinguishable would imply that NGC ~346 is not completely relaxed, and that the majority of the stars would be still close to their birth positions. Also in this case the observed mass segregation would be primordial and would represent a feature of the way the cluster is forming. 

\subsection{Environment and the IMF}
\label{environment}

The Magellanic Clouds differ from the Milky Way in a variety of ways, including their mass densities, tidal fields, chemical abundances, and experiences with recent external perturbations. Both the Magellanic Clouds also are well known for their large populations of relatively massive star clusters: densities of host galaxies and the densities of their star forming sites evidently are not simply related. In the SMC the star formation process on large spatial scales itself differs from that in the Galaxy, more likely being a result of cloud-cloud interactions \citep{zaritsky00,zaritsky04} and the turbulent generation of self-gravitating structures \citep[e.g.][]{elmegreen00}. In some regions this may be further enhanced through the formation of dense shells and supershells which are sites for further star formation \citep{hatzidimitriou05}. Thus the SMC, with its low metallicity and externally driven star formation in a gas--rich system, provides insights into the development of stellar populations in small pre--galactic systems. 

The form of the stellar IMF in the SMC has implications for a wide range of studies. The IMF has been studied for higher mass stars in the SMC: the derived slopes are in good agreement with the values obtained, in similar ranges of masses, in the MW \citep[e.g.][]{massey89,hill94}.  Similarly, existing investigations of the IMF down to 1~M$_{\odot}$ yield Salpeter-like IMF slopes \citep[e.g.][]{mateo88,chiosi07}. Thus while the star formation process might be somehow affected by the conditions of the local interstellar medium, evidently the change in conditions between the SMC and spirals has not significantly affected the IMF.

Most of the stellar mass resides in stars with M$<$1~M$_{\odot}$.  Thus our extension of the lower limit for the PDMF in NGC~346 to 0.8~M$_{\odot}$ in this study adds further support to the overall normalcy of the IMF. Since this result comes from one major star forming complex, our data further suggest that this ``normal'' IMF is imprinted during major formation processes.

On a smaller spatial scale, environment, however, does appear to matter {\em within} the NGC~346 complex.  Consistent with the \citet{sirianni02} work on NGC~330 in the SMC we find that while the overall PDMF in NGC~346 follows a Salpeter form, there are significant local variations. The NGC~346 complex offers, therefore, an ideal opportunity to characterize the occurrence and the nature of the local variations. This will be the objective of future work.

\section{DISCUSSION AND CONCLUSIONS}
\label{conclusions}

Our analysis of the stellar content of NGC~346 indicates that, within all the uncertainties that can affect the determination of a MF, the PDMF slope of NGC 346, at least down to the solar mass regime, is consistent with the value derived by \citet{salpeter55} for the IMF in the solar neighborhood, further confirming that the stellar MF does not strongly depend on the metallicity, or the mass or morphological type of the parent galaxy. 

We studied how the PDMF varies as a function of the distance from the center of NGC~346. We found that the PDMF is quite flat in the center, and becomes steeper in the periphery. As already found in other young star clusters \citep{hillenbrand97,hillenbrand98,sirianni02} the increase in the MF slope with the distance from the center is due to the lack of massive stars in the periphery rather than an excess of low--mass objects there. The projected density of massive stars from the innermost annulus and the outskirts decreases by a factor of 60, whereas low-mass stars are depleted only by a factor of 6; the most massive stars are segregated in the center. 

We conclude that the spatial stellar mass distribution is unlikely to be biased by age effects. SST observations indicate that there is ongoing SF throughout the complex with a lower limit on the total star formation rate of $3.2 \times 10^{-3}\, {\rm M}_\odot\, {\rm yr}^{-1}$ \citep{simon07}. Furthermore \citet{mokiem06} noted a spread of about 5 Myr in the ages of NGC~346 stars, but  did not find any correlation between the age and the position of the stars over the whole cluster. 

S07 noted that stars in NGC~346 appear to be organized in several coeval sub--clusters, embedded in H{\sc ii} gas, and coinciding with the CO clumps analyzed by \citet{rubio00}. YSOs between Class I and Class III are found in 14 of the 16 identified sub--cluster \citep{simon07}. The sub--clusters are also connected by filaments and arcs of gas and dust.  The complex structure of NGC~346 and its young age, compared to its dynamical time, suggest that NGC~346 is not dynamically evolved, and probably not yet relaxed. On the basis of these considerations we conclude that the observed segregation of the most massive stars is likely primordial, and reflects how the cluster formed. 

S07 suggested that in NGC~346 most of the star formation was likely not triggered, but rather resulted from the turbulence driven density variations within a giant interstellar cloud complex, conditions predicted by the hierarchical fragmentation of a turbulent molecular cloud model \citep{elmegreen00,klessen00,bonnell02,bonnell03}. 
According to this model, the fragmentation of the cloud is due to supersonic turbulent motions present in the gas. The turbulence induces the formation of shocks in the gas, and produces filamentary structures \citep{bate03}. The chaotic nature of the turbulence increases locally the density in the filamentary structures. When regions of high density become self--gravitating, they start to collapse to form stars. Simulations show that star formation occurs simultaneously at several different locations in the cloud
\citep{bonnell03}, as appears to be the case for NGC~346.  

\begin{figure}
\epsscale{1.}
\plotone{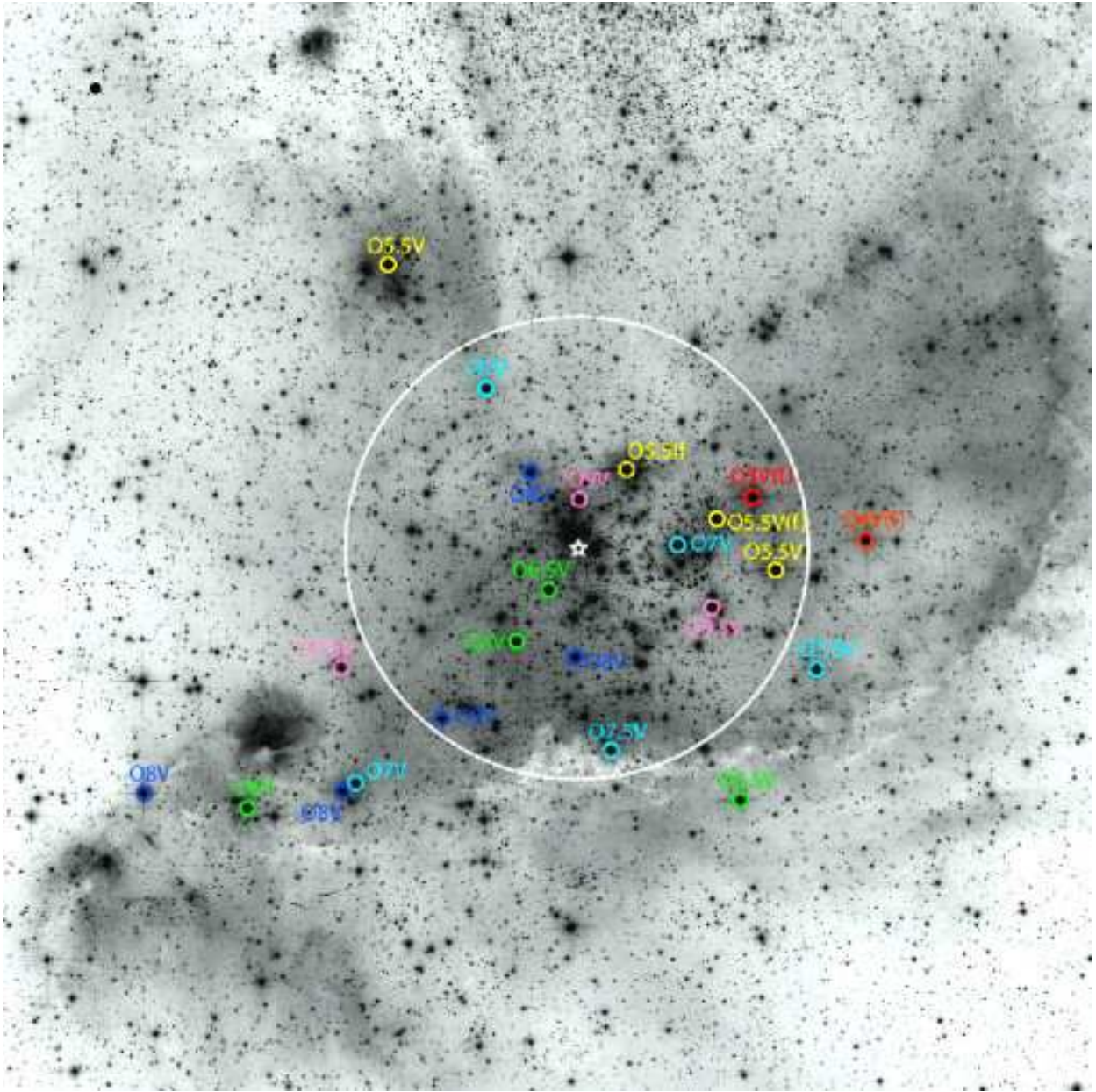}
\caption{\label{f:centro} Spectral classification of the O stars as derived by \citet{massey89}: O3 stars are in red, O4 in orange, O5 in yellow, O6 in green, O7 in cyan, O8 in blue and O9 in purple. The white open star indicates the center of NGC~346, and the white circle the half mass radius, located at 9 pc from the center.}
\end{figure}

In this scenario, stars forming near the cluster center would have higher accretion rates, due to the local higher gas density \citep{larson82,bonnell01}. Simulations by \citet{bonnell01} indicate that although low--mass stars form equally well throughout the entire cluster, the most massive stars are almost exclusively segregated to the central regions. 
It is interesting to note that the two most massive stars (one of them being an O3V spectral type, Figs.~\ref{f:centro}) are $\sim 7\, {\rm pc}$ off the center of the cluster (outside the core of NGC~346, but still within the half mass radius). Did these stars form where they are observed today, in a quite low stellar density region?

\citet{vine03} studied the effect of strong stellar winds and photoionizing fluxes coming from O- and B-type stars on the gas content in a young star cluster. According to their simulations, the gas expulsion is unlikely to remove any initial mass segregation, but it can still result in the ejection of some of the most massive stars to positions outside the cluster core, although still within the half mass radius. According to \citet{bouret03} mechanical power from O-star winds is not a dominant factor in the evolution within N66, but the molecular gas in the original cloud has been strongly photodissociated and photoionized by the O stars \citep{rubio00}.

\citet{niemela86} measured the radial velocity of five of the innermost and brightest stars in NGC~346, and found a velocity dispersion of $\sim 5\, {\rm km\, s}^{-1}$, indicating that, even in $\sim 1\, {\rm Myr}$, they can be formed in the center of the cluster and then ejected where they are observed today. Another possibility is that these stars formed where are observed, as the result of a very efficient competitive accretion process into a potential well \citep{bonnell03}. 

Further spectroscopic analysis of the dynamics of the gas and the sub--clusters can confirm whether the giant molecular cloud which formed NGC~346 underwent under a hierarchical fragmentation. The initial supersonic turbulence originates shocks in the gas which rapidly remove kinetic energy from the gas \citep{ostriker01}: in this case we expect to find low gas velocities around the sub--clusters. A detailed study of the sub--clusters dynamics will also reveal if NGC~346 is or is not already relaxed and virialized. We will present these results in a future paper.

\acknowledgments
We warmly thank Paolo Montegriffo for the software support and Nolan Walborn for helpful suggestions and useful discussions. The photometric conversion table to the ACS filters was kindly provided by Livia Origlia.
Financial support was provided by the Italian MIUR to L. A., and M. T. through PRIN--MIUR--2004 and PRIN--INAF-2005. E. S. has been supported through STScI GO grant GO-1163 and GO-1208.

\clearpage

\begin{table}
\begin{center}
\caption{\label{table} NGC~346 properties derived in this paper.}
\begin{tabular}{lcr}
\tableline
\tableline
NGC~346 MF slope between 1 and 60 $M_\odot$ & & $\Gamma=-1.43\pm0.18$\\
MF slope within 4 pc from the center & & $\Gamma=-1.03\pm0.14$\\
MF slope between 4 and 9 pc from the center & & $\Gamma=-1.36\pm0.09$\\
MF slope between 9 and 14 pc from the center & & $\Gamma=-1.43\pm0.09$\\
MF slope between 14 and 20 pc from the center & & $\Gamma=-2.08\pm0.14$\\
NGC~346 half mass radius & & 9 pc \\
NGC~346 total mass & & $M_\mathrm{Tot}\simeq 3.9\times 10^5\, {\rm M}_\odot$\\
NGC~346 crossing time & & $T_\mathrm{cross}\simeq 1.8\times 10^6\, {\rm yr}$\\
NGC~346 relaxation time && $T_\mathrm{relax}\ge 5.7\times 10^8\, {\rm yr}$\\
NGC~346 mass segregation time && $T_\mathrm{msg}\ge 2\times 10^7\, {\rm yr}$\\
\tableline
\end{tabular}
\end{center}
\end{table}

\clearpage

\clearpage

\end{document}